# WAVES AND CAUSALITY IN HIGHER DIMENSIONS

Paul S. Wesson[1] and James M. Overduin[2,3]

[1] Department of Physics and Astronomy, University of Waterloo, Waterloo, ON, N2L 3G1, Canada.

[2] Department of Physics, Astronomy and Geosciences, Towson University, Towson, MD, 21252, U.S.A.

[3] Department of Physics and Astronomy, Johns Hopkins University, 3400 N. Charles St., Baltimore, MD, 21218, U.S.A.

Abstract: We give a new, wave-like solution of the field equations of five-dimensional relativity. In ordinary three-dimensional space, the waves resemble de Broglie or matter waves, whose puzzling behaviour can be better understood in terms of one or more extra dimensions. Causality is appropriately defined by a null higher-dimensional interval. It may be possible to test the properties of these waves in the laboratory.

# WAVES AND CAUSALITY IN HIGHER DIMENSIONS

1. Introduction

Despite data from the double-slit and related experiments, the theory behind de Broglie or matter waves is not well understood [1-4]. For example, a fundamental feature is that the product of the particle velocity and the wave velocity equals the square of the speed of light, so if the former is subluminal then the latter must be superluminal. This and other puzzles can be better understood if the de Broglie waves observed in ordinary 3D space originate in five or more dimensions [5,6]. The extension of general relativity to five dimensions, as in Membrane theory and Space-Time-Matter theory, is now well established [7]. It is in agreement with extant observations and is widely regarded as a viable step towards a grand-unified theory of all the interactions of physics. In the present work we will present a new, exact solution of the field equations of five-dimensional relativity, and compare this with the approach in four-dimensional spacetime. Our conclusion will be that phenomena involving de Broglie waves may be better understood in terms of the physics of one or more extra dimensions, where causality is defined by setting the extended interval to zero. It might be possible to study such higher-dimensional waves in the laboratory.

2. An Exact 5D Wave Solution

The field equations of five-dimensional relativity are usually defined by the 5D Ricci tensor as $R_{AB} = 0$ ($A, B = 0, 123, 4$ for time, space and the extra coordinate). These 5D equations actually contain Einstein's 4D ones of general relativity, by an old embedding theorem of Campbell. Many exact solutions of the 5D equations are known. But only one exhibits wave-like behaviour of the type shown by experiments on matter waves, and this is restricted by having a constant extra potential [5; 8-10]. It would be of special interest to find a solution which has wave-like properties and involves the extra dimension in a meaningful manner. This because the extra potential represents a scalar field, modulated by spin-0 quanta, which is believed to be of potential importance for both particle physics and cosmology [7]. In this section, we will present such a solution.

Consider the following 5D line element:

$$dS^2 = \exp[\frac{i}{L}(t+ql)]dt^2 - \exp[\frac{2i}{L}(t+\alpha x+ql)]dx^2 - \exp[\frac{2i}{L}(t+\beta y+ql)]dy^2$$
$$- \exp[\frac{2i}{L}(t+\gamma z+ql)]dz^2 - q^2 \exp[\frac{i}{L}(t+ql)]dl^2 \quad . \tag{1}$$

This metric satisfies $R_{AB} = 0$, as may be verified from a tedious calculation by hand or a short run on a computer. It describes a wave in a 5D manifold consisting of ordinary spacetime $(t, xyz)$ plus an extra dimension ($x^4 = l$; we use this symbol to avoid confusion with the Euclidean coordinate). The solution is typified by a constant length $L$, and four dimensionless constants $\alpha, \beta, \gamma$ and $q$ that relate to ordinary 3D space and the extra dimension. All 5 of these constants are arbitrary from the mathematical viewpoint.

The 5D metric (1) is complex, and since this differs from the standard usage in 4D general relativity some comments may be useful. There are actually several solutions of this type in the literature (see ref. 7 for a summary). They usually arise when the field equations possess a symmetry that allows the extra dimension to switch between spacelike and timelike, when one solution is real and the other complex. Both choices of signature are allowed in Space-Time-Matter theory, whereas a timelike extra dimension is the common choice in Membrane theory.

Complex solutions in $N \geq 5D$ typically have properties which cannot be described by real solutions in 4D. They are essential to studying waves and other quantum-related phenomena such as tunneling. The focus here is on wave-like behaviour, so it is natural to consider solutions like (1) above. A discussion of the physics of 5D wave solutions may be found in connection with a previous study [5]. The criterion for the acceptability of any 5D complex solution is that the 4D properties of matter calculated from it should be real. We will find below that the solution (1) satisfies this criterion.

From the physical viewpoint, (1) has several interesting properties which we will illustrate by bringing in the speed of light $c$ and Planck's constant $h$ at appropriate places. Thus the frequency of the wave in (1) is $c/L$, and the wave-numbers in the $x, y$ and $z$ directions are $\alpha/L$, $\beta/L$ and $\gamma/L$. There are actually two dynamical modes of (1) involving the corresponding quantity $q/L$ which is coupled to the extra coordinate $x^4 = l$. When $q$ is part of a complex phase as in (1), the wave number is $q/L$, and the metric has signature $(+---)$ so the extra dimension is spacelike. When $q$ is taken out of the complex phase via $q \rightarrow iq$ in (1), the motion in $l$ is not wavelike but monotonic, and the metric has signature $(+---+)$ so the extra dimension is timelike. Both options are allowed in 5D relativity [7]. We will concentrate on the former case, since we will find that (1) shares several properties with de Broglie waves.

One characteristic property of de Broglie waves is that the product of their phase velocity $v_p$ and group velocity $v_g$ is equal to the square of the speed of light, where the group velocity is identified with the speed of the associated particle [1-4]. The same relation is implied by (1), as may be seen by considering the spacetime part of the wave travelling along the x-axis (say). This is described by $\exp[i(ft + k_x x)]$, where $f$ is the frequency and $k_x$ is the wave number. Here, as noted above, the frequency in conventional units is $f = c/L$. To fix L, we take Planck's law and apply it to the energies of the wave and its associated particle: $E = hf = hc/L = mc^2$, so $L = h/mc$, which is the Compton wavelength of the particle whose mass is m. To fix the wave number $k_x$, we take de Broglie's relation between the wavelength and the momentum of its associated particle, $\lambda_x = h/mv_g$, and invert it to write the wave number as $k_x = (mc/h)(v_g/c) = v_g/cL$. Combining the frequency and the wave number now gives a relation between the phase velocity of the wave and the velocity of the particle:

$$v_p = \frac{f}{k_x} = \frac{c}{L}\left(\frac{cL}{v_g}\right) = \frac{c^2}{v_g} \qquad \text{so} \qquad v_p v_g = c^2 \quad . \tag{2}$$

This is the aforementioned relation for a matter wave and its associated particle. Ordinary particles observed in the laboratory have velocities $v_g < c$ so (2) necessarily implies $v_p > c$. This possibility is clearly present in (1), where the wave number along the x axis is $k_x = \alpha/L$ so the phase velocity is $f/k_x = (c/L)(\alpha/L)^{-1} = c/\alpha$, where $\alpha$ is arbitrary and can be less than unity.

Causality in 5D is most logically defined by the 5D null paths given by $dS^2 = 0$ [5-7]. This includes the conventional 4D paths for both photons and massive particles, given in terms of the 4D interval or proper time by $ds^2 \geq 0$. It has been known for a while that certain 5D metrics admit superluminal velocities, the simplest example being 5D Minkowski space with a timelike extra coordinate. However, such velocities are covered by the condition $dS^2 = 0$, which ensures that all events in the manifold are in causal contact.

The metric (1) shows that motion along (say) the $x$ axis is simple harmonic in nature. If this motion were present in a mechanical system, it would be governed by a 'spring constant' $1/L^2$. The question arises of whether the waves in (1) exist in empty space, or whether they are supported by some kind of fluid. Campbell's theorem, mentioned above, helps to answer this. For it implies that any solution of the apparently empty 5D field equations $R_{AB} = 0$ can be reduced to the 4D Einstein equations $G_{\alpha\beta} = 8\pi T_{\alpha\beta}$, where $T_{\alpha\beta}$ is an *effective* energy-momentum tensor induced by the extra dimension [6, 7: we use geometrical units here]. The precise form of $T_{\alpha\beta}$ for metric (1) can be calculated by longhand or computer, and we have done both. In such problems, the source depends on how the 4D part of the metric is embedded in the extra dimension, and for perfect fluids the density and pressure are typically proportional to $1/\Phi^2$ where $g_{44} \equiv -\Phi^2(x^\gamma, l)$ defines the scalar field. For metric (1), we find that the density and pressure are given by

$$8\pi\rho = \frac{9}{2}\frac{q^2}{L^2\Phi^2} \quad , \quad p = \frac{\rho}{3} \quad . \tag{3}$$

The equation of state is that of radiation or ultra-relativistic particles. The properties of matter depend on the wave number in the extra dimension ($q$) and the magnitude of the scalar field ($\Phi$). They do *not* depend on the wave-numbers of the motion in ordinary 3D space. The oscillations described by the 5D metric (1) are not therefore ordinary electromagnetic waves or conventional gravitational waves (the latter have different properties and propagate through truly empty space). But they exist in 3D space, and share characteristics with a previously-studied case whose background metric describes the Einstein vacuum [5, 6]. In that case, the waves were identified as de Broglie or matter waves. These can exist in any kind of medium, so we believe that the new solution (1) also describes de Broglie waves.

The new solution (1) may be usefully compared with the 5D de Sitter solution, which has been much studied. In 4D, this solution is the basic one with vacuum energy as measured by the cosmological constant, and the field equations have the familiar form $R_{\alpha\beta} = \Lambda g_{\alpha\beta}$. In 5D, the de Sitter solution is known to be not only Ricci-flat with $R_{AB} = 0$ but also Riemann-flat with $R^A{}_{BCD} = 0$. For the new solution, we find that both relations are only satisfied if the constant which appears in the last term of (1) is exactly equal to the square of the wave number ($q$) associated with the extra coordinate. This implies that the existence of the solution (1) and the fact that it is flat in 5D depends, loosely speaking, on the 'size' of the extra dimension.

3. <u>Matter Waves and Causality in 4D and 5D</u>

De Broglie waves as they are understood in 4D raise questions to do with causality [2, 3, 6]. In this section, we wish to give a brief discussion of how matter waves are viewed in 4D spacetime, and then indicate how solutions like (1) above relate to causality in a 5D manifold.

It should be recalled that de Broglie was led to infer that particles have associated waves by essentially comparing the time and space components of the 4-vectors associated with the particle (mass, momentum) and the wave (frequency, wavelength). This procedure leads basically to Planck's law and the definition of what is now called the de Broglie wavelength of a particle. The combination of these relations results in the equation $v_p v_g = c^2$ between the phase velocity of the wave and the group velocity, which latter is identified with the ordinary velocity of the particle (see above). Some workers have been suspicious of the inference that $v_p > c$, since it appears to violate the tenets of special relativity. However, Rindler has argued that the noted relation is indeed valid, using an unusual interpretation of the Lorentz transformations [2]. In the remainder of this section, we will use an alternative and concise method to derive the relation $v_p v_g = c^2$ for matter waves, compare the situation with that for electromagnetic waves, and then discuss implications for causality.

De Broglie's relation between the phase and group velocities of a matter wave may be derived from first principles by using the wave equation and the Minkowski metric. The former is proved in standard texts, and takes the form

$$\frac{\partial^2 \chi}{\partial t^2} = v_p^2 \frac{\partial^2 \chi}{\partial x^2} \quad . \tag{4}$$

Here $\chi$ is a typical property of the medium, such as the potential if the wave travels through a field, and the motion is taken to be along the x-axis at $v_p$, the phase velocity. For a de Broglie wave and its associated particle, it is usual to take $\chi = \exp[i(Et + px)/h]$ where $E$ is the energy, $p$ is the (linear) momentum and h is Planck's constant. Then (4) gives

$$v_p = E/p \quad . \tag{5}$$

In relativistic particle physics, the underlying metric is commonly taken to be the 4D Minkowski one, and the 4-velocities are defined in terms of this and normalized to unity:

$$ds^2 = c^2 dt^2 - dx^2 - dy^2 - dz^2$$

$$u^\alpha u_\alpha \equiv 1, \ u^\alpha \equiv dx^\alpha / ds \quad (x^\alpha = ct, x, y, z).$$

Multiplying this normalization condition throughout by the rest mass $m_0$ and using standard definitions ($E \equiv m_0 c u^0$, $p \equiv m_0 u^1$) there results

$$E^2 = m_0^2 c^4 + p^2 c^2 \quad . \tag{6}$$

This relation has been extensively tested in accelerator experiments. Therefore, the behaviour of the group velocity of the wave (or the velocity of the particle as included in *p*) is well established. However, the phase velocity of the wave is not directly measurable and must be inferred. Using previous relations in (5), the latter reads

$$v_p = E/p = m_0 u^0 / m_0 u^1 = (dt/ds)/(dx/ds) = (dt/dx) = 1/v_g \quad \text{so} \quad v_p v_g = c^2 \quad . \tag{7}$$

This is again de Broglie's relation between the phase velocity of the wave and the group velocity of its associated particle. The same result may be obtained from (5) by expanding (6) in the low-velocity case in terms of $v_g/c \ll 1$, or in the high-velocity case by using Einstein's energy/mass relation. There is no plausible way to avoid the conclusion that particles which can be seen moving at speeds less than *c* should be accompanied by waves which cannot be seen and are moving at speeds greater than *c*.

This result becomes more plausible if we take into account the effects of dispersion in the medium through which the wave propagates. Most media show dispersion at some level, because their microscopic structure causes the phase and group velocities to differ. This is manifested as the spreading of a wave packet, causing the associated particle to become delocalized [1, 3]. Even the vacuum shows behaviour which can be attributed to dispersion [6]. To better understand the implications of the new wave solution (1), it is instructive to consider as a com-parison the propagation of light through a dispersive medium [3]. In that case, the strength of the dispersion, or the difference between $v_g$ and $v_p$, is measured by the variation of the refractive index *n* as a function of the frequency $\omega$ or equivalently the wave-number *k*. Considered as a phenomenological parameter, we would like to evaluate $n(k)$ for de Broglie waves.

The standard relations between the relevant parameters for electromagnetic waves are:

$$\omega(k) = \frac{ck}{n(k)} \tag{8}$$

$$v_p = \frac{\omega(k)}{k} = \frac{c}{n(k)} \tag{9}$$

$$v_g = \frac{c}{[n(\omega) + \omega(dn/d\omega)]} \quad . \tag{10}$$

To these three relations should be added the de Broglie formula (7) which is independent. To obtain $n = n(k)$ from the preceding four relations, it is best to proceed as follows. Multiply (9) and (10) together and use (7) to eliminate *c*, and find

$$\frac{dn}{d\omega} = \frac{(1-n^2)}{n\omega} \quad . \tag{11}$$

Obviously $dn/d\omega$ changes sign at $n=1$, so there are two types of behaviour for $n(\omega)$. Integrating (11) gives

$$n = [1 - (\omega_0/\omega)^2]^{1/2} \quad , \quad n < 1 \quad . \tag{12}$$

The sign inside the square brackets is reversed for $n>1$. The constant $\omega_0$ is mathematically arbitrary, but physically sets a cutoff frequency in devices such as waveguides, where waves with lengths greater than the size of the device cannot propagate. For practical purposes, it is useful to change the variable in (12) to the wave number, using (8). The result is

$$n = [1-(k_0/k)^2]^{-1/2} \quad , \quad n<1 \quad , \tag{13}$$

where as before the sign inside the square brackets is reversed for $n>1$.

In the above account, the refractive index is used as a convenient phenomenological parameter to describe the behaviour of de Broglie waves, and these should not be expected to behave in the same way as electromagnetic waves. In particular, the results (12) and (13) depend on augmenting the optical equations (8)-(10) with de Broglie's equation (7), a procedure which is mathematically valid. However, de Broglie's equation between velocities has special properties, and alters the physics. For example, light waves passing through ordinary materials often show what is defined to be normal dispersion ($dn/d\omega > 0$) for $n>1$, while de Broglie waves in general have this behaviour by (11) for $n<1$. The reason for the differences between light waves and de Broglie waves can be traced to the fact that the latter are related to particles with finite rest mass.

When a particle with finite rest mass travels through any kind of dispersive medium, its associated wave has phase and group velocities which obey the relation $v_p v_g = c^2$ noted before. However, this relation is only one of those which are relevant to the problem (see above). The other two are Planck's law between the energy of a particle and the frequency of its associated wave ($E = mc^2 = hf$), and de Broglie's relation between the wavelength and the momentum of the particle ($\lambda = h/p = h/mv_g$). These three relations are intimately connected in de Broglie's theory of wave mechanics [1, 2, 6]. They form a kind of algebraic triangle, where any two imply the third. Notably, combining Planck's law for the frequency *f* with de Broglie's relation for the wavelength $\lambda$ gives the phase speed $v_p = f\lambda = c^2/v_g$. The velocity relation $v_p v_g = c^2$ is unavoidable, despite its puzzling implications. In fact, insofar as the photoelectric effect (for Planck's law) and wave interference (for de Broglie's relation) are amply tested in the laboratory [8-10], the resulting velocity relation has to be taken seriously. Some workers, though, are reluctant to accept the implication that subluminal particle speeds imply superluminal wave speeds, and argue that the noted relation is not realistic. This because the phase speed $v_p$ is either a pure abstraction, or else irrelevant to the transfer of information.

Both of these objections are suspect, as can be appreciated by the following short thought experiment. Suppose a source *S* of vacuum waves emits in the direction of an observer *O*, and that near *O* there is a particle which is free to move. According to de Broglie's theory, the phase velocity of the waves emitted by *S* affects the ordinary velocity of the particle at *O*. In particular, when *S* emits waves with phase velocity $v_p$, the particle at *O* responds and is seen to move with speed $c^2/v_p$, which can be measured. This process represents at least a crude transfer of influence from *S* to *O*, and an extension of the method to a series of on/off switches can be used in principle to transfer information in binary code, with an effective speed exceeding that of light. It is not difficult to imagine a more sophisticated process, in which the phase speed of the waves is modulated at *S*, representing information which must be recorded with fidelity by the response of the particle at *O*. It should be noted that the spatial separation of *S* and *O* is not really pertinent, even though it can formally approach infinity. This and other aspects of the situation are compatible with the condition $dS^2 = 0$ mentioned above for causality as defined in 5D.

4.  Discussion and Conclusion

In Section 2 we presented a new solution of the 5D field equation whose properties lead to an interpretation in terms of de Broglie or matter waves. In Section 3 we re-examined the status of these waves in 4D, focussing on dispersion and confirming that their properties are indeed unusual compared to other phenomena in spacetime. From these studies, it appears that de Broglie waves are better understood in 5D than 4D.

This conclusion was implied by earlier work [5, 6]. But that work employed a 5D metric quite dissimilar in form to the one used here. This suggests that de Broglie waves may be a common feature of 5D metrics. In this regard, it should be mentioned that material velocities greater than that of light are implied by both modern versions of the Kaluza-Klein approach, namely Space-Time-Matter theory and Membrane theory [7; both approaches can be extended to higher dimensions]. These current theories are in agreement with observations and explain certain aspects of cosmology not covered by general relativity [11-16]. However, the implications of 5D relativity for laboratory-accessible physics have not so far been much investigated.

It appears, after consideration of the physics involved, that a feasible experiment is to measure the phase shift of the eigenstate of a quantum system induced by the extra dimension. This would be technically part of the Berry or geometric phase, and might be detected using approaches developed to investigate the scalar Aharonov-Bohm effect [17-20]. It is hoped to re-port on this and other possible tests in future work. We hope that others will take up the practical side of this subject, since it appears to us that new laboratory experiments may be possible that could detect the influence of an extra dimension.

Acknowledgements